\documentclass[prb,superscriptaddress,twocolumn,amsmath,amssymb]{revtex4}

\usepackage{wasysym}
\usepackage{graphicx}
\usepackage{dcolumn}
\usepackage{bm}

\begin{document}

\renewcommand{\vec}[1]{{\mathbf #1}}
\newcommand{\gvec}[1]{{\boldsymbol #1}}

\title{Nonuniversal dynamic conductance fluctuations
in disordered systems}

\author{N. Cherroret}
\affiliation{
Universit\'{e} Joseph Fourier, Laboratoire de Physique et Mod\'{e}lisation des Milieux Condens\'{e}s, CNRS, 25 rue des Martyrs, BP 166, 38042 Grenoble, France
}
\author{A. \~{P}ena}
\affiliation{
Department of Physics and Astronomy,
University of Texas, San Antonio, TX 78249, USA
}
\author{A.A. Chabanov}
\affiliation{
Department of Physics and Astronomy,
University of Texas, San Antonio, TX 78249, USA
}
\author{S.E. Skipetrov}
\affiliation{
Universit\'{e} Joseph Fourier, Laboratoire de Physique et Mod\'{e}lisation des Milieux Condens\'{e}s, CNRS, 25 rue des Martyrs, BP 166, 38042 Grenoble, France
}

\date{\today}


\begin{abstract}
Sample-to-sample fluctuations of the time-dependent conductance of a system with static disorder have been studied by means of diagrammatic theory and microwave pulsed transmission measurements. The fluctuations of time-dependent conductance are not universal, i.e., depend on sample parameters, in contrast to the universal conductance fluctuations in the steady-state regime. The variance of normalized conductance, determined by the infinite-range intensity correlation $C_3(t)$, is found to increase as a third power of delay time from an exciting pulse, $t$. $C_3(t)$ grows larger than the long-range intensity correlation $C_2(t)$ after a time $t_q \sim \sqrt{\langle g \rangle}\, t_D$ ($t_D$ being the diffusion time, $\langle g \rangle$ being the average dimensionless conductance).
\end{abstract}


\maketitle

\section{Introduction}
\label{intro}

The ability of a conductor to carry electric current is characterized by its conductance $G$ equal to the ratio of the current carried through to the voltage across the conductor. $G$ is finite because of scattering on impurities and lattice defects, which prevents the electrons from propagating ballistically, along straight lines, transforming their trajectories into random walks with a step size $\ell$. The conductance $G$ depends on the exact configuration of impurities inside the conductor. The ensemble average of the conductance, $\langle G \rangle$, can be expressed as $\langle G \rangle = (e^2/h)\langle g \rangle$, where $\langle\ldots\rangle$ represents the average over an ensemble of random impurity configurations, $g$ is the dimensionless conductance, $e$ is the electron charge, and $h$ is the Planck constant. Deviations of conductance from its average value, $\delta G = G - \langle G \rangle$, arise from the interference of electronic wavefunctions and, for this reason, are only important in the mesoscopic regime, i.e., when the electronic wavefunction is phase coherent. A prominent feature of electronic transport in mesoscopic systems is the universal conductance fluctuations (UCF): $\langle\delta G^2\rangle\sim (e^2/h)^2$, for any metal, independent of sample size or disorder strength \cite{akkermans07,webb85,stone85}. The UCF can be expressed in terms of $g$ as $\langle\delta g^2\rangle\sim 1$.

There is a remarkable analogy between electronic transport in mesoscopic systems and wave propagation in random media \cite{feng91,vanrossum99,vanrossum95}. The dimensionless conductance $g$ can be defined for classical waves as the transmittance $T$, that is, the sum over transmission coefficients $T_{ab}$ from an incident mode $a$ to an outgoing mode $b$, at a given frequency $\nu$, $g\equiv T = \Sigma_{ab}T_{ab}$. Here the modes $a$ and $b$ correspond to transverse eigenmodes of the empty waveguide and form a complete set of basis functions, sufficient to represent any wave field entering or leaving the waveguide. A given mode $a$ is characterized by a transverse wave vector $\vec{q}_a$. The statement of UCF is then equivalent to $\Sigma_{aba'b'}\langle\delta T_{ab}\,\delta T_{a'b'}\rangle\sim 1$. The origin of UCF can be traced to correlations of transmission coefficients $T_{ab}$ of a random sample. Feng, Kane, Lee, and Stone \cite{stone85} have expressed the intensity correlation function, $C_{aba'b'}=\langle\delta T_{ab}\,\delta T_{a'b'}\rangle$, as a series using $1/\langle g \rangle$ as the expansion parameter: $C_{aba'b'}=C_1+C_2+C_3$.
$C_1 \sim \langle T_{ab} \rangle^2$ is the largest contribution to $C_{aba'b'}$ but it is different from zero only when $\Delta \vec{q}_a = \Delta \vec{q}_b$, where $\Delta \vec{q}_a = \vec{q}_a - \vec{q}_{a'}$ and similarly for $\Delta \vec{q}_b$. It is therefore ``short range'' in both $\Delta \vec{q}_a$ (at a given $\Delta \vec{q}_b$) and $\Delta \vec{q}_b$ (at a given $\Delta \vec{q}_a$). Its contribution to $\Sigma_{aba'b'}\langle\delta T_{ab}\,\delta T_{a'b'}\rangle$ is negligible.
$C_2 \sim \langle T_{ab} \rangle^2/\langle g \rangle$ is
$\langle g \rangle$ times smaller, but it decays only algebraically with $\Delta \vec{q}_a$ and $\Delta \vec{q}_b$. For this reason, $C_2$ is often referred to as a ``long-range'' correlation function. Its contribution to $\langle\delta T^2\rangle$ is still much less than unity. It is $C_3 \sim \langle T_{ab} \rangle^2/\langle g \rangle^2$  ---
the smallest of the three correlation functions and independent of $a$, $a'$, $b$ and $b'$ --- that explains UCF: ${\rm var}(T)\sim\Sigma_{aba'b'}C_3\sim 1$.
Because $C_3$ does not decay neither with $\Delta \vec{q}_a$ nor with $\Delta \vec{q}_b$, it is reffered to as an ``infinite-range'' correlation function.

UCF have been extensively studied in mesoscopic systems both experimentally \cite{webb85} and theoretically \cite{stone85} (see Refs.\ \onlinecite{akkermans07} and \onlinecite{vanrossum99} for reviews). They have also been demonstrated in random media, with light \cite{scheffold98} and microwaves \cite{tregoures04}, through the observation of the infinite-range ($C_3$) intensity correlation. However, all studies of UCF yet conducted have only concerned the steady-state conductance.
In this paper, we study fluctuations of the time-dependent conductance of a disordered system by means of diagrammatic theory and microwave pulsed transmission measurements. We show that the universality of conductance fluctuations is lost in the dynamic experiment: the fluctuations depend on the length of the sample and the mean free path $\ell$. The variance of normalized transmittance is observed to increase as a third power of delay time from an exciting pulse. A comparative analysis of the underlying mesoscopic correlations allows us to identify a new characteristic time scale $t_q = \sqrt{\langle g \rangle}\, t_D$, $t_D$ being the diffusion time, at which $C_3(t)$ grows larger than $C_2(t)$.

\newpage
\section{Theory}
\label{theory}

In order to study the dynamic conductance fluctuations, we consider a pulsed wave propagating through a quasi-1D waveguide with static disorder, of cross section $A$ and length $L \gg \ell$, parallel to the $z$-axis. This geometry also corresponds to the theoretical model of a disordered mesoscopic conductor. A pulsed excitation $\psi_0(t)$ at a point $\gvec{\rho}_0$ on the input surface $A_{\rm in}$ of the waveguide ($z = 0$) results in a complex intensity distribution $I(\gvec{\rho}, t; \gvec{\rho}_0)$ at the output surface $A_{\rm out}$ of the waveguide ($z = L$). To avoid confusion, we acknowledge the source position $\gvec{\rho}_0$ as an explicit argument of $I$. With a proper normalization of $\psi_0(t)$, the transmittance $T(t)$ is given by
\begin{eqnarray}
T(t) = \int_{A_{\rm out}}\!\!\! d^2 \gvec{\rho} \int_{A_{\rm in}}\!\!\! d^2 \gvec{\rho_0}
I(\gvec{\rho}, t; \gvec{\rho}_0),
\label{t}
\end{eqnarray}
and its variance is
\begin{eqnarray}
\langle{\delta T(t)^2}\rangle &=& \int_{A_{\rm out}}\!\!\! d^2 \gvec{\rho} \int_{A_{\rm in}}\!\!\! d^2 \gvec{\rho_0}
\int_{A_{\rm out}}\!\!\! d^2 \gvec{\rho}^{\prime} \int_{A_{\rm in}}\!\!\! d^2 \gvec{\rho_0}^{\prime}
\nonumber \\
&\times&C(\gvec{\rho}, \gvec{\rho}_0; \gvec{\rho}^{\prime}, \gvec{\rho}_0^{\prime}; t),
\label{dg2}
\end{eqnarray}
where $C(\gvec{\rho}, \gvec{\rho}_0; \gvec{\rho}^{\prime}, \gvec{\rho}_0^{\prime}; t)=\langle\delta I(\gvec{\rho}, t; \gvec{\rho}_0) \delta I(\gvec{\rho}^{\prime}, t; \gvec{\rho}_0^{\prime})\rangle$ is the time-dependent spatial correlation function of the intensity fluctuations, $\delta I(\gvec{\rho}, t; \gvec{\rho}_0) = I(\gvec{\rho}, t; \gvec{\rho}_0) - \langle I(\gvec{\rho}, t; \gvec{\rho}_0)\rangle$.

\begin{figure}[t]
\includegraphics[width=\columnwidth]{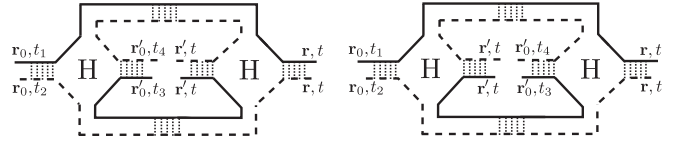}
\caption{\label{Diagram}
Two of the four diagrams contributing to the kernel $K$ of Eq.\ (\ref{kernel}). The solid and dashed lines represent the Green's function ${\cal G}$ and its complex conjugate ${\cal G}^*$, respectively. Dotted lines denote scattering events. $H$ is the Hikami box \cite{gorkov79,hikami81}, which describes an exchange of partners between 4 Green's functions.}
\end{figure}

In order to calculate $C$, we express the transmitted field $\psi(t)$ at a point $\vec{r} = \{\gvec{\rho}, L \}$ in terms of the exciting field $\psi_0(t')$ at $\vec{r}_0 = \{\gvec{\rho_0}, 0 \}$ with the help of the Green's function ${\cal G}$ of Helmholtz equation:
\begin{eqnarray}
\psi(\vec{r}, t; \vec{r}_0) =
\int_{-\infty}^t d t^{\prime} {\cal G}(\vec{r}, t; \vec{r}_0, t^{\prime})
\psi_0(t^{\prime}).
\label{psi}
\end{eqnarray}
Using the definition of intensity, $I(\gvec{\rho}, t; \gvec{\rho}_0) \equiv | \psi(\vec{r}, t; \vec{r}_0)|^2$, we obtain
\begin{eqnarray}
C(\gvec{\rho}, \gvec{\rho}_0; \gvec{\rho}^{\prime}, \gvec{\rho}_0^{\prime}; t) &=&
\int_{-\infty}^t \prod_{i=1}^4 d t_i
K(\textbf{r}, \textbf{r}_0, \textbf{r}^{\prime}, \textbf{r}_0^{\prime}; t, \{t_i\})
\nonumber \\
&\times&\psi_0(t_1) \psi_0^*(t_2) \psi_0(t_3) \psi_0^*(t_4),
\label{cc}
\end{eqnarray}
where the kernel $K$ is given by the connected part of a product of four Green's functions, averaged over disorder:
\begin{eqnarray}
K &=&
\langle{\cal G}(\vec{r}, t; \vec{r}_0, t_1) {\cal G}^*(\vec{r}, t; \vec{r}_0, t_2)
\nonumber \\
&\times& {\cal G}(\vec{r}^{\prime}, t; \vec{r}_0^{\prime}, t_3)
{\cal G}^*(\vec{r}^{\prime}, t; \vec{r}_0^{\prime}, t_4)\rangle
\nonumber \\
&-&  \langle{{\cal G}(\vec{r}, t; \vec{r}_0, t_1) {\cal G}^*(\vec{r}, t; \vec{r}_0, t_2)}\rangle
\nonumber \\
&\times&
\langle{{\cal G}(\vec{r}^{\prime}, t; \vec{r}_0^{\prime}, t_3)
{\cal G}^*(\vec{r}^{\prime}, t; \vec{r}_0^{\prime}, t_4)}\rangle.
\label{kernel}
\end{eqnarray}
It is important to note that $K$ can be written in the Fourier space as a correlation function of ${\cal G}$'s at \emph{four} different frequencies. As a consequence, our calculation does not reduce to that of Ref.\ \onlinecite{vanrossum95} where the correlation function of steady-state conductances at \emph{two} different frequencies was studied.
Among many diagrams contributing to $K$ we are interested in those which do not vanish when both $\left| \gvec{\rho} - \gvec{\rho}^{\prime} \right|$ and $\left| \gvec{\rho}_0 - \gvec{\rho}_0^{\prime} \right|$ exceed $\lambda_0$, the wavelength in the medium at the carrier frequency $\nu_0$. For a waveguide with a large number of transverse channels, $N \gg 1$, such diagrams give the leading-order contribution to $\langle{\delta T(t)^2}\rangle$. Similar to the steady-state case \cite{akkermans07,vanrossum99,stone85}, the diagrams of interest contain two pairs of Green's functions forming two ladders and exchanging partners twice on the way from $\vec{r}_0$ and $\vec{r_0}^{\prime}$ to $\vec{r}$ and $\vec{r}^{\prime}$. We have identified four such diagrams, two of them are shown in Fig.\ \ref{Diagram}. They are similar to those encountered in the steady-state case \cite{akkermans07,vanrossum99,stone85}, but include time-dependent Green's functions.

When the diagrams for $K$ are calculated, they are substituted into Eq.\ (\ref{cc}), which is used to calculate $\langle{\delta T(t)^2}\rangle$. At this point it is necessary to assume a specific shape of the pulsed excitation. Here we consider a Gaussian pulse, $|\psi_0(t)|^2 \sim\exp(-t^2/t_p^2)$, and study the variance of the normalized transmittance $s(t)\equiv T(t)/\langle T(t) \rangle$,
\begin{eqnarray}
\mathrm{var}[s(t)] = \frac{\langle{\delta T(t)^2}\rangle}{\langle{T(t)}\rangle^{2}},
\label{dcf}
\end{eqnarray}
for various pulse durations $t_p$ \footnote{When analyzing the dynamic conductance fluctuations, it might appear natural to consider the limit of infinitely short pulses, $t_p \rightarrow 0$. We shall see, however, that the time-dependent conductance becomes self-averaging and its variance vanishes in this limit.}.

For pulse duration $t_p$ significantly greater than the diffusion time, $t_p\gg t_D=(L + 2 z_0)^2/\pi^2 D$, where $D$ is the diffusion coefficient, $z_0\sim\ell$ is the extrapolation length, and for short time delays from the center of the exciting pulse, $t \ll t_p$, we recover the steady-state result, $\mathrm{var}[s(t)]= 2/15{\langle g \rangle}^{2}$. A particularly simple result is obtained for short pulses, $t_p \ll t_D$, and long time delays, $t \gg t_D$, when the average transmittance takes the form $\langle{T(t)}\rangle \propto \exp(-t/t_D)$. Only the diagrams of Fig.\ \ref{Diagram} contribute to $\mathrm{var}[s(t)]$ in the leading order in $t_p/t_D$. Keeping only terms that do not decay exponentially with $t/t_D$, we obtain \begin{eqnarray}
\mathrm{var}[s(t)] = \frac{2}{15 {\langle g\rangle}^{2}} \times
\frac{5 \sqrt{2}}{\pi^3 \sqrt{\pi}} \frac{t_p}{t_D}
\left(  \frac{t}{t_D} \right)^3.
\label{c3t}
\end{eqnarray}
This result is independent of absorption, if the pulse duration $t_p$ is much less than the absorption time $t_a$, $t_p \ll t_a$. This condition can be readily met in practice, opening interesting perspectives for probing scattering parameters of random media without necessarily knowing absorption inside the medium. For long pulses, $t_p \gtrsim t_D$, all the four above mentioned diagrams contribute to $\langle{\delta T(t)^2}\rangle$. At long times $t \gg t_p$, $\mathrm{var}[s(t)]$ then takes the form of a cubic polynomial with the same coefficient in front of $t^3$ as in Eq. (\ref{c3t}) and coefficients in front of lower powers of $t$ that have to be calculated numerically and will be presented elsewhere.

To derive Eq.\ (\ref{c3t}), we explicitly made use of the exponential decay of the average transmittance with time: $\langle{T(t)}\rangle \propto \exp(-t/t_D)$ for $t \gg t_D$. Meanwhile, it is known that $\langle{T(t)}\rangle$ exhibits deviations from a simple exponential at long times $t > t_q = \sqrt{\langle g \rangle} t_D$ (weak localization corrections, see Refs.\ \onlinecite{chabanov03,skip04a,cheung04}). However, these deviations are of order $1/\langle g \rangle$ for $\langle g \rangle \gg 1$ and, therefore, taking them into account would yield corrections only of order $1/\langle g \rangle^3$ or higher in Eq.\ (\ref{c3t}). In the present paper, we restrict ourselves to the calculation of $\mathrm{var}[s(t)]$ in the leading order in $1/\langle g \rangle \ll 1$ and hence drop all higher-order terms. In addition,
$\mathrm{var}[s(t)] = \langle{\delta T(t)^2}\rangle/\langle{T(t)}\rangle^{2}$ and corrections of similar type exist for the numerator $\langle{\delta T(t)^2}\rangle$ as well as for the denominator $\langle{T(t)}\rangle^{2}$ of this ratio, leading to a non-exponential decay of both of them with time. As a result, a partial cancelation of these corrections might be expected in the final result.

As is evident from Eq.\ (\ref{c3t}), the fluctuations of the time-dependent conductance are not universal, in contrast to the steady-state case. They explicitly depend on the mean free path, sample length, and other parameters of the disordered sample. According to Eq.\ (\ref{c3t}), at long times $\mathrm{var}[s(t)]$ can reach values well in excess of the steady-state result $\mathrm{var}(s) = 2/15 {\langle g\rangle}^2$. Other consequences of Eq.\ (\ref{c3t}) are the cubic growth of conductance fluctuations with time and their direct proportionality to the pulse duration $t_p$, implying that $T(t)$ is a self-averaging quantity in the limit of $t_p \rightarrow 0$.

\section{Experimental setup and sample characterization}
\label{exp}

Because the measurement of transmittance $T$ requires the summation of $T_{ab}$ over all incoming and outgoing modes (or, equivalently, integration over input and output surfaces of the waveguide), it represents a real challenge. In our microwave experiment, we make use of the relation between fluctuations of $T$ and the intensity correlation function $C_3$ discussed in the Introduction. Because for vector waves, the notion of ``mode'' includes the polarization state as well, it is easy to show that if $\vec{e}_a \perp \vec{e}_{a'}$ and $\vec{e}_b \perp \vec{e}_{b'}$, the only contribution to the correlation function $\langle \delta T_{ab} \delta T_{a'b'} \rangle$ comes from $C_3$ term \cite{tregoures04}:
$\langle \delta T_{ab}(t) \delta T_{a'b'}(t) \rangle = C_3(t)$. Here $\vec{e}_a$ is the unit vector describing the linear polarization state of the mode $a$. A link between $\mathrm{var}[s(t)]$ and $C_3(t)$ is then readily found: $\mathrm{var}[s(t)]=C_3(t)/\langle T_{ab}(t)\rangle^2$. Vector nature of electromagnetic waves leads to doubling of the number of transverse channels $N$ as compared to the scalar case, and hence to doubling of $\langle g \rangle$ in Eq.\ (\ref{c3t}).

\begin{figure}[t]
\includegraphics[width=0.95\columnwidth]{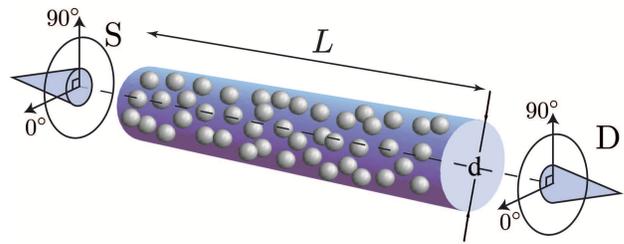}
\caption{\label{Setup}
(color online). Drawing of the experimental setup. Two microwave horns, S (source) and D (detector), are positioned in front of and behind the random sample. The polarizations of the incident and detected waves can be rotated by rotating the horns about their axes ($z$-axis).}
\end{figure}

Measurements of the correlation function of transmission coefficients $\langle \delta T_{ab}(t) \delta T_{a'b'}(t) \rangle$ for $\vec{e}_a \perp \vec{e}_{a'}$ and $\vec{e}_b \perp \vec{e}_{b'}$ have been made for microwave radiation transmitted through random mixtures of alumina spheres. Alumina spheres with diameter 0.64 cm and index of refraction 3.14 are embedded within Styrofoam shells to produce a sample with alumina volume fraction 0.03. The sample is contained within a copper tube with diameter of 4.4 cm and plastic end pieces. Linearly polarized microwave radiation is launched and detected by conical horns placed 20 cm in front of and behind the sample, as shown in Fig. \ref{Setup}. Microwave field spectra are taken for cross-polarization orientations of both the source and the detector for 15 000 configurations produced by briefly rotating the tube, in samples of length 61, 76.2, and 91.4 cm. Measurements are made over the frequency range 18.7--19.7 GHz, away from sphere resonances and far from the localization threshold. At 19.2 GHz, the number of transverse channels is $N=54$, with the effective refractive index of the sample, $n_{\rm eff}=1.175$. The transport mean free path estimated from Mie theory is $\ell =3.76$ cm, giving $\langle g\rangle = 4N\ell/3(L+2z_0) \simeq 4.10$, 3.33, and 2.81 for $L=61$, 76.2, and 91.4 cm, respectively (we set $z_0 = 2\ell/3$). The temporal response to a Gaussian pulse of width $t_p$ peaked at $t=0$ is obtained by taking the Fourier transform of the field spectrum multiplied by a Gaussian envelope of width $(2\pi t_p)^{-1}$ centered at $\nu_0=19.2$ GHz. The field of the time response is squared to give the transmitted intensity $I(t)$.

It is worthwhile to note that our experimental setup is similar to that of Ref. \onlinecite{chabanov04}, except that the cross-polarized source and detector horns select the $C_3$ correlation function, whereas the measurements of Ref. \onlinecite{chabanov04} were dominated by $C_1$ and $C_2$. Indeed, one can define three distinct transmission quantities: intensity transmission coefficient $T_{ab}$, total transmission $T_a = \sum_b T_{ab}$ and transmittance $T = \sum_{ab} T_{ab}$. Fluctuations of $T_{ab}$ are described by $C_1$ correlation function, whereas $C_2$ and $C_3$ govern fluctuations of $T_a$ and $T$, respectively \cite{akkermans07,vanrossum99}. Fluctuations of $T_{ab}$ and $T_a$ were studied in Ref.\ \onlinecite{chabanov04}, whereas the focus of the present paper is on fluctuations of $T$. It is important to understand that because $T_a$ is obtained from $T_{ab}$ by summing only over outgoing modes $b$, whereas $T$ includes the summation over incoming modes $a$ as well, fluctuations of $T_a$ and $T$ are dominated by different scattering processes.

\begin{figure}[t]
\includegraphics[width=\columnwidth]{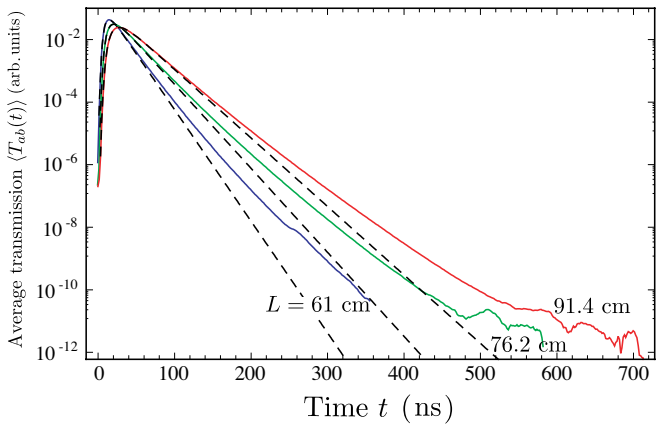}
\caption{\label{avint}
(color online). Time-dependent average transmission coefficient (in arbitrary units) of disordered microwave waveguides of three different lengths $L = 61$, $76.2$ and $91.4$ cm (solid lines). Polarizations of incident and detected modes $a$ and $b$ do not have any importance for this measurement because multiply scattered waves are completely depolarized. Dashed lines are theoretical fits to the data at short times (up to twice the peak arrival time) using diffusion approximation \cite{Alfano90}. Diffusion times $t_D$ extracted from the fits are $t_D = 17.5$, $27.3$ and $39.6$ ns, respectively.}
\end{figure}

To characterize our samples, we first estimate the absorption time $t_a$. To this end, we measure the decay rate of transmission in a sample of length $L = 25.4$ cm with copper end caps, weakly coupled to the measurement port (see Ref.\ \onlinecite{zhang09} for details of this measurement). We obtain $t_a=39.7$ ns. We then study the average transmission coefficient $\langle T_{ab}(t) \rangle$. A fit of diffusion theory \cite{Alfano90} to $\langle T_{ab}(t) \rangle$ at short times (up to twice the peak arrival time of the pulse) yields diffusion times $t_D=17.5$, 27.3, and 39.6 ns for samples of lengths 61, 76.2, and 91.4 cm, respectively (see Fig.\ \ref{avint}).
The fits in Fig.\ \ref{avint} exhibit notable deviations from the data at long times. This is expected from the previous work \cite{chabanov03,skip04a,cheung04} and is due to weak localization effects in our strongly disordered samples (see discussion in Sec.\ \ref{theory}).

\section{Results and discussion}
\label{res}

\begin{figure}[t]
\includegraphics[width=\columnwidth]{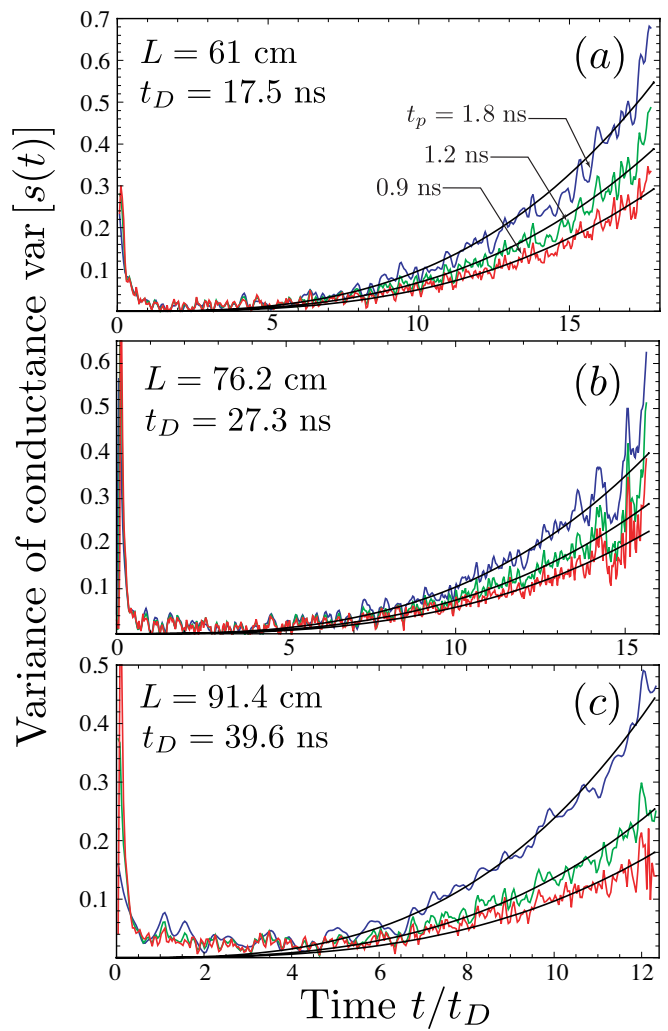}
\caption{\label{exp}
(color online). Variance of normalized transmittance, $\mathrm{var}[s(t)]$, in samples of three different lengths $L$ for three different pulse durations $t_p$ (curves in the same order for the three panels). The diffusion time $t_D$ extracted from the fit of diffusion theory to the average transmission coefficient $\langle T_{ab}(t) \rangle$ (see Fig.\ \ref{avint}) is provided for each sample. Smooth lines plotted through the data points are the fitting curves of Eq.\ (\ref{c3t}) with $\langle g \rangle = 4.17$(a), 3.21(b) and 1.85(c). Only the data for $t\geq 3 t_D$ were used for the fitting.}
\end{figure}

The variance of normalized transmittance, $\mathrm{var}[s(t)]$, for the pulse durations $t_p = 0.9$, 1.2, and 1.8 ns for each of the samples studied is shown in Fig. \ref{exp}. The experimental data are fit with Eq.\ (\ref{c3t}) using $\langle g\rangle$ as the only fitting parameter and imposing that $\langle g\rangle$ is the same for the three pulse durations for given $L$. The values of $\langle g \rangle$ found from the fit are 4.17, 3.21 and 1.85 for samples of lengths 61, 76.2, and 91.4 cm, respectively. These values differ slightly from the estimates, except for the longest sample. Going beyond the ``two $H$-box'' diagrams of Fig.\ \ref{Diagram} [and hence, including higher-order terms in $1/\langle g \rangle$ in Eq.\ (\ref{c3t})] is necessary for a better quantitative agreement with the data for samples with small values of $\langle g \rangle$.

\begin{figure}[t]
\includegraphics[width=\columnwidth]{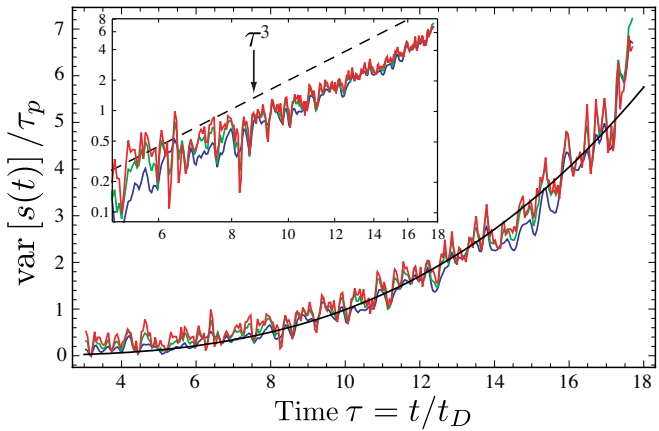}
\caption{\label{scaling}
(color online). Scaling properties of $\mathrm{var}[s(t)] \propto \tau_p \tau^3$, with $\tau_p = t_p/t_D$ and $\tau=t/t_D$, at long times, $t \gg t_D$, are demonstrated with the data for $L = 61$ cm and $t_p$'s as in Fig.\ \ref{exp}(a). The theoretical (smooth) curve is for $\langle g \rangle = 4.17$. The inset shows the same plot on a log-log scale.}
\end{figure}

To demonstrate that Eq.\ (\ref{c3t}) correctly captures the key scaling properties of $\mathrm{var}[s(t)]$ for $t \gg t_D$, we introduce the dimensionless time $\tau = t/t_D$ and the normalized pulse duration $\tau_p = t_p/t_D$, and plot $\mathrm{var}[s(t)] / \tau_p$ versus $\tau$ for $L = 61$ cm in Fig.\ \ref{scaling}. The three curves corresponding to the three different pulse durations fall on a single curve, demonstrating $\mathrm{var}[s(t)]\propto\tau_p$. The same data plotted on a log-log scale (inset of Fig.\ \ref{scaling}) show that $\mathrm{var}[s(t)]\propto\tau^3$. Thus, $\mathrm{var}[s(t)] \propto \tau_p \tau^3$, as predicted by Eq.\ (\ref{c3t}).

The power-law growth of $\mathrm{var}[s(t)]$ with time can be qualitatively understood from the following reasoning. In the path picture of wave propagation and according to the diagrams of Fig.\ \ref{Diagram}, the value of $\mathrm{var}[s(t)]$ is proportional to the probability for two wave paths of length $ct$ to cross twice inside the sample. The probability of a single crossing is $p(t) \sim ct/NL$ \cite{akkermans07,skip04}. The probability that the two paths cross first during the time interval (0, $t^{\prime}$) and then a second time during the time interval ($t^{\prime}$, $t$) is $p(t^{\prime}) p(t - t^{\prime})$. $\mathrm{Var}[s(t)] \propto t^3/\langle g \rangle^2$ is obtained by integrating over $t^{\prime}$ from $0$ to $t$.

Keeping in mind the relation between $\mathrm{var}[s(t)]$ and $C_3(t)$, we can use Eq.\ (\ref{c3t}) to compare $C_3(t)$ with the long-range intensity correlation function $C_2(t) \propto (1/\langle g \rangle) (t_p/t_D) (t/t_D)$, previously studied in Ref.\ \onlinecite{skip04}. Such a comparison suggests an important time scale, $t_q =  \sqrt{\langle g \rangle}\, t_D$, intermediate between the Thouless time $t_{Th}\sim t_D$ and the Heisenberg time $t_H \sim \langle g \rangle t_D$. This time scale appeared previously in the context of weak localization in classically chaotic \cite{casati97} and disordered \cite{skip04a,mirlin00} systems, but its role in the analysis of \emph{fluctuations} of transport properties has never been identified. Because for short pulses in the long-time limit we can write $C_2(t) \sim (t_p/t_q) (t/t_q)$ and $C_3(t) \sim (t_p/t_q) (t/t_q)^3$, the formally next-order contribution (in $1/\langle g \rangle$ expansion) to the intensity correlation function, $C_3(t)$ becomes larger than $C_2(t)$ when $t > t_q$. Even though this might indicate that some new physics come into play at such long times, $\mathrm{var}[s(t)]$ of Eq. (\ref{c3t}) agrees well with the experimental data for $t > t_q$, and even for $t > t_H$, when the discrete mode structure of the disordered sample is expected to play a role \cite{mirlin00}.

\section{Conclusion}
\label{concl}

In conclusion, we presented a theoretical and experimental study of fluctuations of the dynamic conductance, describing the response of a system with static disorder to a pulsed excitation. The variance of normalized conductance is not universal and increases with delay time from an exciting pulse, $t$, reaching values well in excess of the steady-state variance. A comparative analysis of the underlying mesoscopic correlations $C_3(t)$ and $C_2(t)$ allowed us to identify a new characteristic time scale $t_q$, being intermediate between the Thouless and Heisenberg times, and to explain its role in dynamic statistics of transport. Our results apply to both electronic transport in mesoscopic systems and wave propagation in random media.

\acknowledgements
We thank B.A. van Tiggelen for fruitful discussions and P. Krueger for help in constructing the experimental apparatus. S.E.S. acknowledges financial support of the French ANR (Project No. 06-BLAN-0096 CAROL) and the French Ministry of Education and Research.

\end{document}